\documentclass[twocolumn,showpacs,preprintnumbers]{revtex4}
\usepackage{amssymb}
\usepackage{amsfonts}
\usepackage{amsmath}
\usepackage{graphicx}
\usepackage{dcolumn}
\usepackage{bm}

\begin{document}

\title{Topological oscillations of the magnetoconductance in disordered GaAs
layers.}
\author{S. S. Murzin$^{\text{1,4}}$, A. G. M. Jansen$^{\text{2,4}}$, and I.
Claus$^{\text{3,4}}$ }
\affiliation{$^{\text{1}}$Institute of Solid State Physics RAS, 142432, Chernogolovka,
Moscow District., Russia\\
$^{\text{2}}$Service de Physique Statistique, Magn\'{e}tisme, et
Supraconductivit\'{e}, D\'{e}partement de Recherche \\
Fondamentale sur la Mati\`{e}re Condens\'{e}e, CEA-Grenoble, 38054 Grenoble
Cedex 9, France\\
$^{\text{3}}$Center for Nonlinear Phenomena and Complex Systems, Facult\'{e}
des Sciences, Universit\'{e} \\
Libre de Bruxelles, Campus Plaine, Code Postal 231, B-1050 Brussels, Belgium%
\\
$^{\text{4}}$Grenoble High Magnetic Field Laboratory, Max-Planck-Institut f%
\"{u}r Festk\"{o}rperforschung and Centre National de la Recherche
Scientifique, BP 166, F-38042, Grenoble Cedex 9, France}

\begin{abstract}
Oscillatory variations of the diagonal ($G_{xx}$) and Hall ($G_{xy}$)
magnetoconductances are discussed in view of topological scaling effects
giving rise to the quantum Hall effect. They occur in a field range without
oscillations of the density of states due to Landau quantization, and are,
therefore, totally different from the Shubnikov-de Haas oscillations. Such
oscillations are experimentally observed in disordered GaAs layers in the
extreme quantum limit of applied magnetic field with a good description by
the unified scaling theory of the integer and fractional quantum Hall effect.
\end{abstract}

\pacs{73.50.Jt; 73.61.Ey; 73.40.Hm}
\maketitle

The integer quantum Hall effect (QHE) is usually observed at high magnetic
fields, $\omega _{c}\tau \gg 1$ ($\omega _{c}=eB/m$ is the cyclotron
frequency, $\tau $ is the transport relaxation time), and its appearance
develops from the Shubnikov-de Haas oscillations based on the Landau
quantization of the two-dimensional (2D) electron system. However, the
scaling treatment of the integer QHE \cite{Pr83} predicts the existence of
the QHE without the Landau quantization of the electron spectrum. It could
exist even at low magnetic field $\omega _{c}\tau \ll 1$ \cite{Khm} in the
absence of magnetoquantum oscillations of the density of states. The QHE at
low magnetic fields $\omega _{c}\tau \ll 1$ has not been observed so far,
probably because extremely low temperatures are required \cite{Huck}. In
addition, the QHE could exist in a layer whose thickness $d$ is much larger
than the electron transport mean free path $l$, i.e. $d\gg l$, in the
extreme quantum limit (EQL) of applied magnetic field, where only the lowest
Landau level is occupied. Such a layer has a three-dimensional (3D) "bare"
(non-renormalized) electron spectrum without oscillations of the density of
states in the EQL. In this situation the QHE has been observed in heavily
Si-doped n-type GaAs layers \cite{MJL,MCJ}.

Here, we address the problem of the arising of the QHE in the absence of
magnetoquantum oscillations of the density of states. In this case the
variation with temperature of the diagonal conductance per square ($G_{xx}$)
and Hall conductance ($G_{xy}$) is due to diffusive interference effects
(below $G_{xx}$ and $G_{xy}$ are taken in units $e^{2}/h$), which in a
scaling approach can be described by the renormalization-group equations.
For comparison, according to the conventional theory, the temperature
dependence of the Shubnikov-de Haas oscillations preceding the QHE is due to
thermal broadening of the Fermi distribution \cite{Ando}. At the moment, two
theories give explicit expressions for the renormalization-group equations.
The first theory has been derived for both integer and fractional QHE and
for any value of $G_{xx}$ \cite{Dolan}. It is based on the assumption that a
certain symmetry group unifies the structure of the integer and fractional
quantum Hall states \cite{Lut,Dolan,Bur}. This so-called unified scaling
(US) theory describes well the shape of the scaling flow diagram depicting
the coupled evolution of $G_{xx}$ and $G_{xy}$ for decreasing temperatures
in heavily Si-doped n-type GaAs layers with different thickness for a wide
range of $G_{xx}$ values \cite{fl}. The second theory has been developed in
the\ "dilute instanton gas" approximation (DIGA), firstly for
non-interacting \cite{Pr87} and then for interacting electrons \cite{PB95}.
Both theories are developed for a totally spin-polarized electron system.
For $2\pi G_{xx}\gg 1$ they predict an oscillating topological term in the
scaling $\beta $-function with the same periodicity. However, they differ in
predictions on the oscillation amplitude. The oscillating topological term
in the $\beta $-function should lead to oscillations in the magnetic-field
dependence of $G_{xx}$ and $G_{xy}$ which are not related to oscillations in
the density of states like, e.g., for the case of the Shubnikov-de Haas
oscillations.

In the presented work we derive explicit expressions for the topological
oscillations of the Hall conductivity $G_{xy}$ for both theories, and
compare them with experiment for thick ($d\gg l$) disordered heavily
Si-doped GaAs layers with rather large $G_{xx}$ and $G_{xy}$ compared to
unity. The layers studied before in Ref.\cite{MJL,MCJ} have a 3D "bare"
electron spectrum. However, below ~4~K the characteristic diffusion lengths, 
$L_{\varphi }=(D_{zz}\tau _{\varphi })^{1/2}$ and $L_{T}=(D_{zz}\hbar
/k_{B}T)^{1/2}$, for coherent diffusive transport increase to values larger
than $d$, and the system becomes 2D for coherent diffusive phenomena ($%
D_{zz} $ is the diffusion coefficient of electrons along the magnetic field, 
$\tau _{\varphi }$ is the phase breaking time).

The US theory describes the renormalization group flow of the conductances
by the equation \cite{Dolan}, 
\begin{equation}
s-s_{0}=-\ln \left( f/f_{0}\right) ,  \label{s1}
\end{equation}%
for a real parameter $s$ monotonically depending on temperature, where $%
G\equiv G_{xy}+iG_{xx}$, $f_{0}=f(s_{0})$ and 
\begin{equation}
f=-\frac{\left( \sum_{n=-\infty }^{\infty }q^{n^{2}}\right) ^{4}\left(
\sum_{n=-\infty }^{\infty }(-1)^{n}q^{n^{2}}\right) ^{4}}{\left(
2\sum_{n=0}^{\infty }q^{\left( n+1/2\right) ^{2}}\right) ^{8}},
\end{equation}%
with $q=\exp (i\pi G)$. For $|q|^{2}=\exp (-2\pi G_{xx})\ll 1$, the function 
$f=-1/(256q^{2})+3/32+O(q^{2})$ and Eq.(\ref{s1}) is reduced to 
\begin{equation}
s-s_{0}\approx i2\pi \left( G-G^{0}\right) +24\left( e^{i2\pi G}-e^{i2\pi
G^{0}}\right)  \label{s2}
\end{equation}%
In the first-order approximation by ignoring the last oscillating term in
Eq.(\ref{s2}), this equation has the solution 
\begin{equation}
G_{xx}^{1}=G_{xx}^{0}-(s-s_{0})/2\pi ,\qquad G_{xy}^{1}=G_{xy}^{0}.
\end{equation}%
In the second-order approximation, the solution looks like 
\begin{eqnarray}
G_{xx} &=&G_{xx}^{1}+\frac{12}{\pi }\left[ e^{-2\pi G_{xx}^{1}}-e^{-2\pi
G_{xx}^{0}}\right] \cos (2\pi G_{xy}^{0})~~~~~  \label{O1} \\
G_{xy} &=&G_{xy}^{0}-\frac{12}{\pi }\left[ e^{-2\pi G_{xx}^{1}}-e^{-2\pi
G_{xx}^{0}}\right] \sin (2\pi G_{xy}^{0}).~~~~~  \label{O2}
\end{eqnarray}%
This is a solution of Eq.(\ref{s2}) for fixed $s$. However, for our
experiment we are interested in the solution for fixed temperature $T$. In
the first-order approximation it should coincide with the result of the
first-order perturbation theory for the electron-electron interaction in
coherent diffusive transport leading to logarithmic temperature-dependent
corrections in the diagonal conductance 
\begin{equation}
G_{xx}^{T}=G_{xx}^{0}+\lambda /2\pi \,\ln \left( T/T_{0}\right) ,
\label{lnT}
\end{equation}%
without any temperature dependence in the Hall conductance \cite{AA}.
Therefore, $s=-\lambda \ln (T)$ in this approximation. For totally
spin-polarized electron system $\lambda =1$ \cite{Fin}.

In second order, $s$ will oscillate as a function of $G_{xy}^{0}$ at fixed
temperature $T$ and will give additional oscillating term in Eq.(\ref{O1}),
but the relation between $s$ and $T$ is unknown and the amplitude of the $%
G_{xx}$ oscillations can not be found. In this respect we note, that the
last term in Eq.(\ref{O1}) shows maxima at integer $G_{xy}^{0}$ as opposed
to the expected minima for the integer QHE. The difference between $%
G_{xx}^{1}$ and $G_{xx}^{T}$ can be ignored in the exponents of Eq.(\ref{O2}%
). Therefore the Hall conductivity $G_{xy}$ oscillates as a function of the
"bare" Hall conductance $G_{xy}^{0}$ and hence as a function of the magnetic
field $B$, with amplitude 
\begin{eqnarray}
A_{xy}^{US} &=&\frac{12}{\pi }\left[ e^{-2\pi G_{xx}^{T}}-e^{-2\pi
G_{xx}^{0}}\right]  \label{DT} \\
&=&\frac{12}{\pi }e^{-2\pi G_{xx}^{0}}\left[ \left( T_{0}/T\right) ^{\lambda
}-1\right] ,  \notag
\end{eqnarray}%
as found by substituting $G_{xx}^{T}$ (Eq.(\ref{lnT})) for $G_{xx}^{1}$ in
Eq.(\ref{O2}). This dependence is totally different from the exponential
variation with temperature of the Shubnikov-de Haas oscillations.

In the "dilute instanton gas" approximation for the case of interacting
electrons \cite{PB95} 
\begin{eqnarray}
\frac{dG_{xx}}{d\ln L} &=&-\frac{\lambda }{\pi }-D_{1}G_{xx}^{2}e^{-2\pi
G_{xx}}\cos (2\pi G_{xy})~~~~~  \label{P1} \\
\frac{dG_{xy}}{d\ln L} &=&-D_{1}G_{xx}^{2}e^{-2\pi G_{xx}}\sin (2\pi
G_{xy}).~~~~~  \label{P2}
\end{eqnarray}%
Here $L\approx (\hbar D_{xx}/k_{B}T)^{1/2}$ and $D_{1}=64\pi /e\approx 74.0$%
. Solving the quotient of these equations by ignoring terms of order $\exp
(-4\pi G_{xx})$ one obtains 
\begin{equation}
G_{xy}=G_{xy}^{0}-\frac{\pi D_{1}}{\lambda }\left[
F(G_{xx}^{T})-F(G_{xx}^{0})\right] \sin (2\pi G_{xy}^{0})~~~~~
\end{equation}%
where $F(x)=1/4\pi ^{3}\,\left( 2\pi ^{2}x^{2}+2\pi x+1\right) \exp (-2\pi
x) $.

Both theories have been developed for a totally spin-polarized electron
system. However, in a real system electrons can have two different spin
projections. For the case of non-interacting electrons, the electrons can be
described in terms of two independent, totally spin polarized systems in the
absence of spin-flip scattering. This approach remains valid for interacting
electrons as well, if the triplet part of the constant of interaction is
much smaller than the singlet one \cite{AA,Fin}, because only the
interaction between electrons with the same spin leads to a renormalization
of the conductance in this case. For the small spin-splitting in strongly
disordered GaAs, the conductances of the electron systems with different
spin projection ($G_{ij}^{\uparrow }$ and $G_{ij}^{\downarrow }$) are
approximately equal to half the measured conductance, i.e. $G_{ij}^{\uparrow
}\approx G_{ij}^{\downarrow }\approx G_{ij}/2$. It allows us to compare
quantitatively the experimental results with the theories. For large
spin-splitting this is impossible, because $G_{ij}^{\uparrow }$ and $%
G_{ij}^{\downarrow }$ are different, and only the sum $G_{ij}^{\uparrow
}+G_{ij}^{\downarrow }$ can be measured.

\begin{figure}[tbp]
\includegraphics[width=8cm,clip]{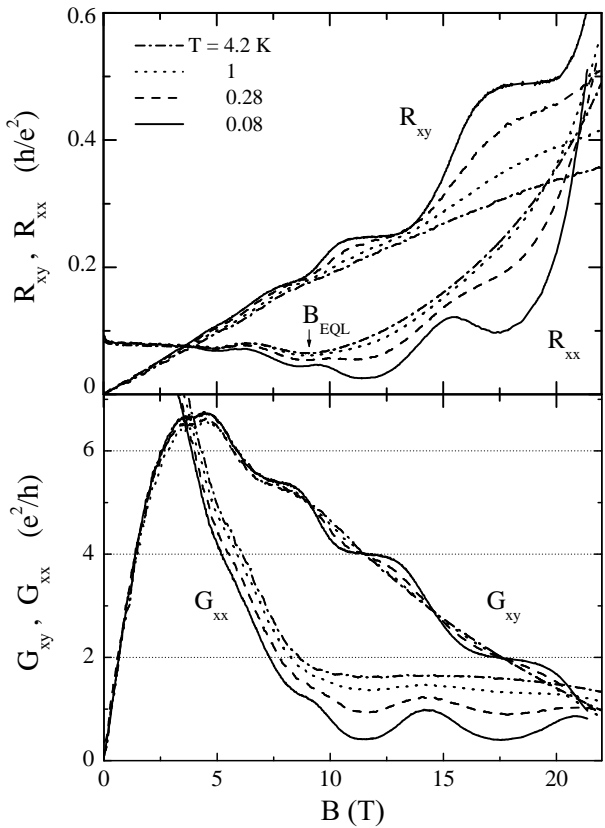}
\caption{Magnetic field dependence of the diagonal ($R_{xx}$, per square)
and Hall ($R_{xy}$) resistance and of the diagonal ($G_{xx}$) and Hall ($%
G_{xy}$) conductance for sample~2 in a magnetic field perpendicular to the
heavily doped GaAs layer (thickness 100~nm) at different temperatures. The
arrow indicates the field $B_{EQL}$ of the extreme quantum limit.}
\label{RG}
\end{figure}

\begin{figure}[tbp]
\includegraphics[width=7cm,clip]{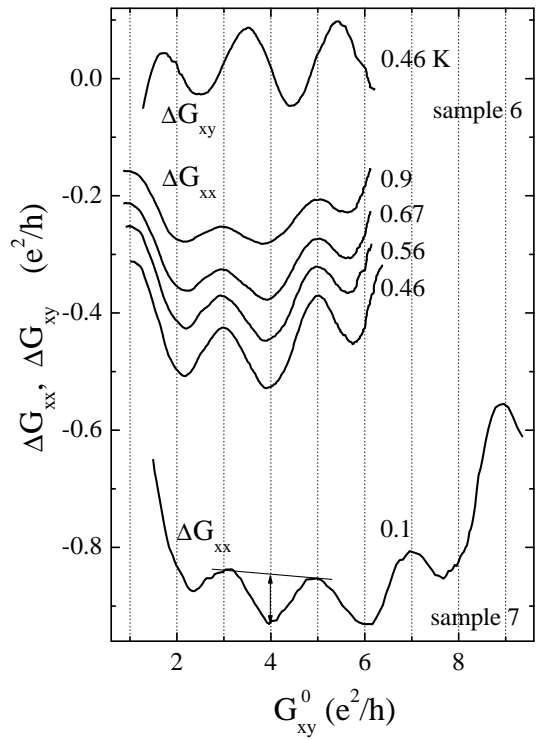}
\caption{Residual variation for the diagonal $\Delta G_{xx}$ and for Hall
conductance $\Delta G_{xy}$ after subtraction of the 4.2-K values at
different temperatures, for samples 6 and 7. Numbers near curves indicate
temperatures in K.}
\label{osc}
\end{figure}

The investigated heavily Si-doped n-type GaAs layers sandwiched between
undoped GaAs were prepared by molecular-beam epitaxy. The nominal thickness $%
d$ equals 100~nm for the layers 2, 3, 6, and 140~nm for layer 7. The
Si-donor bulk concentration $n$ equals 1.8, 2.5, 1.6, and $3\times 10^{17}$
cm$^{-3}$\ for samples 2, 3, 6, and 7 as derived from the period of the
Shubnikov-de Haas oscillations at $B<5$~T. The mobilities of the samples at $%
T=4.2$ K are 2400, 2500, 2600 and $2600$~cm$^{2}$/Vs, and the electron
densities per square $N_{s}$ as derived from the slope of the Hall
resistance $R_{xy}$ in weak magnetic fields ($0.5-3$~T) at $T=4.2$~K are
1.26, 2, 2.08 and $2.86\times 10^{12}$ cm$^{-2}$ for samples 2, 3, 6 and 7,
respectively. For all samples the electron transport mean free path $l$ is
around 30~nm at zero magnetic field. The detailed structure of the samples
is described in Ref.\cite{MJL}.

In Fig.\ref{RG} the magnetotransport data of the diagonal ($R_{xx}$, per
square) and Hall ($R_{xy}$) resistance (both given in units of $h/e^{2}$),
and of the diagonal ($G_{xx}$) and Hall ($G_{xy}$) conductance are plotted
for sample~2. At 4.2~K, the magnetoresistance shows the typical behavior of
bulk material with weak Shubnikov-de Haas oscillations for increasing field $%
B$ and a strong monotonous upturn in the extreme quantum limit (EQL) where
only the lowest Landau level is occupied. At lower temperatures $R_{xy}$, $%
R_{xx}$,$\ G_{xy}$, and $G_{xx}$ start to oscillate. Minima of $G_{xx}$ and
of $\left\vert \partial G_{xy}/\partial B\right\vert $ arise at magnetic
fields where $G_{xy}$ at 4 K attains even-integer values, in accordance with
both theories mentioned above. These oscillatory structures develop into the
QHE at the lowest temperatures where $R_{xy}$ and $G_{xy}$ reveal remarkable
steps near the values $R_{xy}=1/2$ and 1/4, and $G_{xx}=2$, and 4. In the
corresponding fields pronounced minima are observed in $R_{xx}$ and $G_{xx}$%
. Note that, contrary to the QHE structures, the amplitude of the weak
Shubnikov-de Haas oscillations below the EQL does not depend on temperature
because the thermal damping factor $2\pi ^{2}k_{B}T/[\hbar \omega _{c}\sinh
(2\pi ^{2}k_{B}T/\hbar \omega _{c})]=0.994$ is close to 1 for $B=5$ T at $T=1
$ K. Similar but less pronounced structures are observed for the other
samples investigated. Moreover, for samples 3 and 7 additional minima of $%
G_{xx}$ and of $\left\vert \partial G_{xy}/\partial B\right\vert $ are
observed, at fields where $G_{xy}=6$ at $T=4$~K.

The size quantization could result in oscillatory structures in the
magnetotransport data in the EQL in a pure layer with ballistic motion
across the layer when $l/d\gg 1$. In our case, however, $l/d\approx 0.2\div
0.3$ in zero magnetic field, which ratio even decreases in the EQL for the
mean free path along the field. The 3D character of the "bare" electron
spectrum of the samples has been confirmed in experiments in a tilted
magnetic field \cite{MCJ}. Note that the absence of oscillations at $T=4.2$%
~K can not be explained by temperature broadening of the oscillatory
structures, because disorder broadening dominates largely with $\hbar /\tau
\gg k_{B}T$ (for our samples $\hbar /\tau k_{B}> 80$~K).

\begin{figure}[t]
\includegraphics[width=8cm,clip]{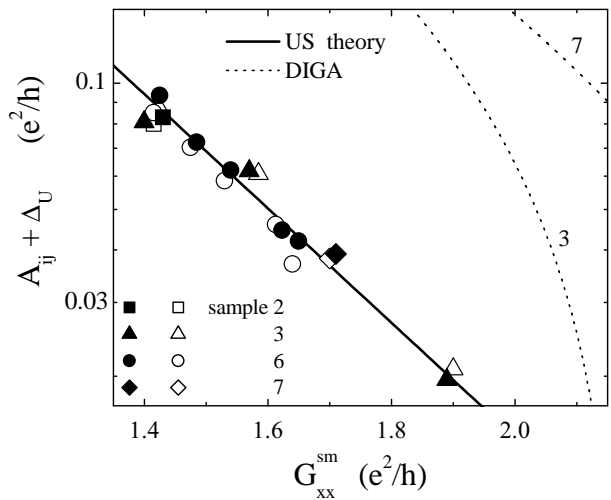}
\caption{Amplitude $A_{ij}$ of the topological oscillation of the Hall
(solid symbols) and diagonal (open symbols) conductance plus $\Delta
_{U}\equiv 24/\protect\pi \exp (-\protect\pi G_{xx}^{0})$ as a function of
the smooth part $G_{xx}^{sm}$ of the diagonal conductance for four samples.
The full line shows the dependence $24/\protect\pi \exp \left( -\protect\pi %
G_{xx}^{sm}\right) $ following from the unified scaling theory. The dotted
lines show the result of the "dilute instanton gas" approximation theory for
samples 3 and 7.}
\label{A}
\end{figure}

In Fig.\ref{osc} we plot the residual variation $\Delta
G_{xx}(T)=G_{xx}(T)-G_{xx}^{0}$ as a function of $G_{xy}^{0}$\ for sample 6
at different temperatures, $\Delta G_{xy}=G_{xy}(T)-G_{xy}^{0}$ at $T=0.46$\
K for sample 6, and $\Delta G_{xx}$ at $T=0.1$\ K for the thickest sample 7.
Here $G_{ij}^{0}$ is the conductance at $T=4.2$ K taken as the "bare"
conductance (see below). Both $\Delta G_{xx}$ and $\Delta G_{xy}$\ oscillate
with comparable amplitudes under the same conditions of applied field and
temperature. The minima of $\Delta G_{xx}$ are at even integer values of $%
G_{xy}^{0}$ (slightly shifted in case of a superimposed smooth variation of $%
\Delta G_{xx}$) and the minima of $\Delta G_{xy}$ are shifted on $+0.5$ unit
in the $G_{xy}^{0}$ scale, in accordance with theory \cite{Dolan}.

The smoothly varying part of $G_{xx}$, by ignoring the oscillatory part,
decreases for decreasing temperature while that of $G_{xy}$ does not change.
The temperature dependence of the smooth part $G_{xx}^{sm}$ of the diagonal
conductance, taken as the midpoint value of the arrow in Fig.\ref{osc}, is
well described by the first-order electron-electron-interaction correction
(Eq.(\ref{lnT})) with $\lambda =1.9$ for samples 2, 3, and 6, and $\lambda =2
$ for sample 7 in the temperature range from 0.15 to 1~K followed by a
saturation around 4.2~K. These values are close to the theoretical upper
limit $\lambda =2$ for a system with two spins \cite{AA,Fin}, corresponding
to a negligibly small triplet part of the electron-electron interaction. The
choice of the 4.2~K value for the "bare" conductance $G_{xx}^{0}$ agrees
with the saturation of $G_{xx}^{sm}$ around $T=4.2$~K.

The amplitudes $A_{ij}$ of the oscillations of $G_{xx}$, and $G_{xy}$,
conductances are very similar as shown in Fig.\ref{A} where the sum $%
A_{ij}+\Delta _{U}$ is plotted as a function of the smooth part of the
diagonal conductance $G_{xx}^{sm}$ for all our samples with $\Delta
_{U}=24/\pi \exp (-\pi G_{xx}^{0})$. The values of $\Delta _{U}=0.044$,
0.009, 0.02, and 0.002 for samples 2, 3, 6, and 7, respectively, are smaller
than the corresponding values of $A_{ij}$. The experimental data are rather
well described by the result of the US theory for $A_{xy}$ (Eq. \ref{DT})
applied to the total conductance of two independent electron systems of
opposite spin. Although showing a very similar dependence, $A_{xx}$ can not
be deduced in frame of this theory. The DIGA theory predicts much larger
amplitudes than experimentally observed, as shown by the dotted lines in Fig.%
\ref{A} for $A_{xy}^{DIGA}+\Delta _{U}$ according DIGA theory for samples 3
and 7.

In summary, due to topological scaling effects oscillations of the diagonal
and Hall magnetoconductances can exist when there are no oscillations of the
density of states due to Landau quantization. The oscillations observed in
the extreme quantum limit of the applied magnetic field in disordered GaAs
layers, with thickness larger than the electron transport mean free path,
fall into this category. The oscillations of $G_{xy}$ are quantitatively
well described by the unified scaling theory for the integer and fractional
quantum Hall effect \cite{Dolan}. Their amplitude is much smaller than the
"dilute instanton gas" approximation \cite{PB95} predicts.

We would like to thank I. S. Burmistrov for helpful discussions, and N. T.
Moshegov, A. I. Toropov, K. Eberl, and B. Lemke for their help in the
preparation of the samples. This work is supported by RFBR and INTAS.


\begin{thebibliography}{99}
\bibitem{Pr83} H. Levine, S. B. Libby, and A. M. M. Pruisken, Phys.\ Rev.\
Lett.\ \textbf{51}, 1915 (1983); A. M. M. Pruisken in \emph{The Quantum Hall
Effect}, edited by R. E. Prange and S. M. Girven, Springer-Verlag, 1990.

\bibitem{Khm} D. E. Khmel'nitski\u{\i}, Pis'ma Zh.\ Eksp.\ Teor.\ Fiz.\ 
\textbf{38}, 454 (1983) [JETP Lett. \textbf{38}, 552 (1983)]; Phys. Lett. A 
\textbf{106}, 182 (1984).

\bibitem{Huck} Bodo Huckestein, Phys.\ Rev.\ Lett.\ \textbf{84}, 3141 (2000).

\bibitem{MJL} S. S Murzin, A. G. M. Jansen, and P. v. d. Linden, Phys.\
Rev.\ Lett.\ \textbf{80}, 2681 (1998); S.S. Murzin, I. Claus, A.G.M. Jansen,
N.T. Moshegov, A.I. Toropov, and K. Eberl, Phys. Rev. B \textbf{59}, 7330
(1999). S. S. Murzin, M. Weiss, A. G. M. Jansen and K. Eberl, Phys. Rev. B 
\textbf{64}, 233309 (2001).

\bibitem{MCJ} S. S. Murzin, I. Claus, and A. G. M. Jansen, Pis'ma Zh.\
Eksp.\ Teor.\ Fiz. \textbf{68}, 305 (1998) [JETP Lett. \textbf{68}, 327
(1998)].

\bibitem{Ando} T. Ando, A. B. Fowler, and F.Stern, Rev. Mod. Phys. \textbf{54%
}, 437 (1982).

\bibitem{Dolan} B. P. Dolan, Nucl. Phys. B \textbf{554}[FS], 487 (1999);
cond-mat/9809294.

\bibitem{Lut} C.A. L\"{u}tken and G.G. Ross, Phys.\ Rev.\ B \textbf{45},
11837 (1992); \textbf{48}, 2500 (1993).

\bibitem{Bur} C. P. Burgess and B. P. Dolan, Phys.\ Rev.\ B \textbf{63},
155309 (2001).

\bibitem{fl} S. S. Murzin, M. Weiss, A. G. M. Jansen and K. Eberl, Phys.
Rev. B \textbf{66}, 233314 (2002).

\bibitem{Pr87} A. M. M. Pruisken, Nucl. Phys. B \textbf{285}[FS19], 719
(1987); Nucl. Phys. B \textbf{290}[FS20], 61 (1987).

\bibitem{PB95} A.M.M. Pruisken and M.A. Baranov, Europhys. Lett. \textbf{31}%
, 543 (1995).

\bibitem{AA} B. L. Al'tshuler and A. G. Aronov, in \emph{Electron-Electron
Interaction in Disordered Systems}, edited by A. L. Efros and M. Pollak,
North-Holland, Amsterdam, 1987.

\bibitem{Fin} A. M. Finkelstein, Zh. Eksp. Teor. Fiz. \textbf{86}, 367
(1984) [Sov. Phys. JETP \textbf{59}, 212 (1984)].
\end{thebibliography}
\end{document}